 \documentclass[final,1p,times]{elsarticle}

\usepackage{amssymb}

\usepackage{lineno}

\journal{Astroparticle Physics}

\begin{document}

\begin{frontmatter}



\title{Cosmic Ray Studies with the Fermi Gamma-ray Space Telescope Large Area Telescope}


\author[djt]{D. J. Thompson}
\author[lb]{L Baldini}
\author[yu]{Y. Uchiyama}
\address[djt]{NASA Goddard Space Flight Center, Greenbelt, MD 20771 USA}
\address[lb]{Istituto Nazionale di Fisica Nucleare, Sezione di Pisa, I-56127 Pisa, Italy}
\address[yu]{SLAC National Accelerator Laboratory, 2575 Sand Hill Road M/S 29,
Menlo Park, CA 94025 USA}


\begin{abstract}
The Large Area Telescope (LAT) on the {\it Fermi Gamma-ray Space Telescope} provides both direct and indirect measurements of Galactic cosmic rays (CR).  The LAT high-statistics observations of the 7 GeV - 1 TeV electron plus positron spectrum and limits on spatial anisotropy constrain models for this cosmic-ray component.  On a Galactic scale, the LAT observations indicate that cosmic-ray sources may be more plentiful in the outer Galaxy than expected or that the scale height of the cosmic-ray diffusive halo is larger than conventional models. 
Production of cosmic rays in supernova remnants (SNR) is supported by the LAT $\gamma$-ray studies of several of these, both young SNR and those interacting with molecular clouds.  
\end{abstract}


\end{frontmatter}


\section{Introduction}
\label{}
Among the earliest stimuli for studying cosmic $\gamma$ rays was the recognition by Hayakawa~\citet{1952_Hayakawa} and Morrison~\citet{1958_Morrison} that cosmic ray particles interacting with interstellar matter would inevitably produce high-energy $\gamma$ rays.  The first high-energy $\gamma$-ray sky map by OSO-3~\citep{1973_OSO-3} confirmed this expectation.  Gamma-ray and cosmic-ray studies, both observational and theoretical, have been closely linked ever since, with principal results for photon energies above 30 MeV coming from satellite detectors (SAS-2~\cite{1975_SAS-2}, COS-B~\cite{1989_COS-B},  EGRET/Compton Gamma Ray Observatory~\cite{2008_EGRET}) and ground-based TeV telescopes (for a recent review, see~\cite{2009_TeVREview}).

While sharing many experimental techniques, cosmic-ray and $\gamma$-ray studies offer complementary approaches.  Cosmic-ray detectors measure the particles directly in the solar neighborhood, including detailed spectral and composition analysis.  Observations with $\gamma$-ray telescopes study cosmic-ray interactions at a distance, providing indirect insight about the particles.  The $\gamma$-ray observations have the potential to trace sources of cosmic rays, since the photons are not deflected by Galactic magnetic fields as the charged particles are. 

The present review concentrates on results from the Large Area Telescope on the {\it Fermi Gamma-ray Space Telescope} satellite, which was launched on 11 June 2008.  The {\it Fermi} LAT offers three lines of research related to cosmic rays:
\begin{enumerate}
\item Because the LAT is a particle detector, it can make direct measurements of cosmic rays, especially cosmic-ray electrons (CRE). 
\item The large-scale $\gamma$-ray properties of our own and other galaxies offer some insight into the distribution of cosmic rays and their sources. 
\item The LAT measures the $\gamma$-ray properties of likely acceleration/interaction sites of cosmic rays such as supernova remnants.
\end{enumerate}

\section{The {\it Fermi} Large Area Telescope}

The LAT is a $\gamma$-ray telescope~\citep{LAT09_instrument}, designed to distinguish photons in the energy range 20 MeV to more than 300 GeV from the high background of energetic charged particles at the 565 km altitude orbit of the {\it Fermi} satellite. The tracking section has 36 layers of silicon strip detectors to record the tracks of charged particles, interleaved with 16 layers of tungsten foil (12 thin layers, 0.03 radiation length, at the top or {\it Front} of the instrument, followed by 4 thick layers, 0.18 radiation length, in the {\it Back} section) to promote $\gamma$-ray conversion into electron-positron pairs.
Beneath the tracker is a calorimeter comprised of an 8-layer array of CsI crystals (1.08 radiation length per layer) to determine the 
$\gamma$-ray energy. The tracker is surrounded by segmented charged-particle anticoincidence detectors (ACD; plastic scintillators 
with photomultiplier tubes) to reject cosmic-ray background events. The LAT's high sensitivity 
 stems from a large peak effective area ($\sim$8000~cm$^2$), wide field of view ($\sim$2.4~sr), good background rejection, superior angular resolution (68\% containment angle $\sim$0.6$^\circ$ at 1~GeV for the {\it Front} section and about a factor of 2 larger for the {\it Back} section), and excellent observing efficiency (keeping the sky in the field of view with scanning observations).  In scanning mode, the LAT observes the entire sky every 2 orbits (about 3 hours).  Pre-launch predictions of the instrument performance are described in~\cite{LAT09_instrument}. 

In addition to being a $\gamma$-ray telescope, the LAT is
intrinsically an electron detector. One of the biggest analysis challenges for electron studies
is the separation of CRE from the much more abundant
hadrons (primarily protons). Key to the background rejection is
the LAT capability to discriminate between electromagnetic and hadronic
showers, based on the different event topology in the three subsystems.
This is largely in common with the photon analysis, the most prominent
difference being the use of the ACD.

Unlike previous generations of pair conversion telescopes,
the LAT basically triggers on all the charged particles crossing the active
volume, with no harsh photon selection built into the hardware
trigger.
This drastic change of approach is largely due to the use of silicon
detectors, allowing precise tracking with essentially no detector-induced dead
time. Some event filtering is necessary prior to downlink in order
to fit the data volume into the allocated telemetry bandwidth. The onboard
filter is really tailored to preserve the $\gamma$-ray efficiency by discarding events that are unlikely to be $\gamma$ rays. 
The rate of events depositing more than 20~GeV into the calorimeter, however, is small
enough that we can afford effectively to disengage the filter above that energy, transmitting to the ground
 high-energy charged particle events along with the $\gamma$ rays.

\section{Direct Cosmic Ray Measurements}

A century after the pioneering cosmic-ray studies by Domenico~Pacini and Victor~Francis Hess
we have a huge body of knowledge about these energetic particles. In the LAT energy range,
many instruments have measured the energy spectra of CR particles and their composition, including not only protons and electrons, but also heavier nuclei. They can be roughly divided into
\emph{inclusive} measurements (i.e. $p$ and $e^- + e^+$ spectra) and
measurements of the charge composition (i.e. $\bar{p}/(p + \bar{p})$ and
$e^+/(e^- + e^+)$), providing complementary information on the origin,
acceleration and propagation of cosmic rays. This dichotomy has a parallel in
the variety of experimental techniques routinely exploited since the mid-1960's,
in that cosmic-ray instruments generally fall in one of the two categories:
calorimetric experiments (that cannot distinguish, per se, the charge sign)
and magnetic spectrometers.

\begin{figure}[htbp!]
  \begin{center}
    \includegraphics[width=\textwidth]{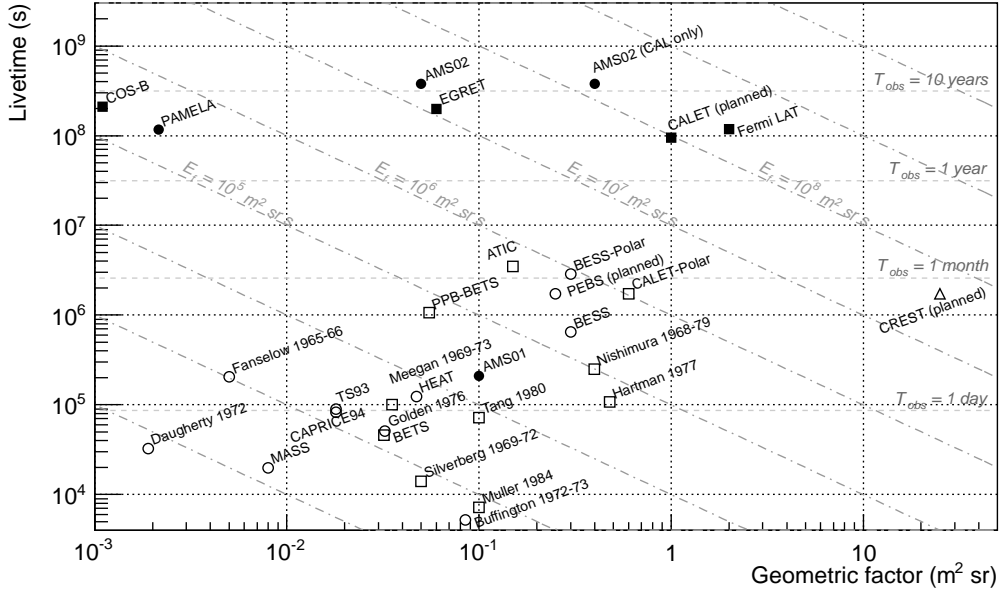}
  \end{center}
  \caption{Compilation of the characteristics (in terms of geometric factor
    and collection time) of some balloon and satellite cosmic-ray and $\gamma$-ray detectors.
    Care has been taken in extracting the relevant 
    information from the original
    publications in a consistent fashion. We believe that the information in
    the plot is accurate within a factor of $\approx 2$, which is adequate
    for the illustrative purpose of the plot itself.
    Filled markers indicate space-based experiments, while open markers
    indicate balloon experiments; conversely squares indicate calorimetric
    experiments, while circles indicate magnetic spectrometers.}
  \label{fig:cr_experiments}
\end{figure}

Figure~\ref{fig:cr_experiments} is a partial compilation of such
instruments, classified by acceptance%
\footnote{The acceptance, or geometry factor ($G_{\rm F}$), is the integral of
  the effective area over the instrument field of view and constitutes the
  proportionality factor between the event rate in the detector and
  the incident flux.}
and effective observation time. It is not a trivial task to give an exhaustive
and unbiased account of the huge experimental effort that turned the first
observations into precision science. In many cases the geometry factor is energy
dependent, and some modern instruments (particularly the magnetic spectrometers)
can operate in different configurations with different acceptances. In addition, each detector is sensitive in a different energy band (for any given
observable), which makes a fair comparison extremely difficult. Despite these caveats, Figure~\ref{fig:cr_experiments} is accurate enough to illustrate
some interesting aspects of the historical development of cosmic-ray
measurements: calorimetric experiments feature, in general, a larger acceptance
than magnetic spectrometers; and space experiments have the obvious advantage
of a relatively longer collection time. Figure~\ref{fig:cr_experiments} 
also  puts {\it Fermi} into the context of this development: the LAT's unique
combination of large acceptance and long observation time results in an
unrivaled exposure factor%
\footnote{The exposure factor is the product of the geometry factor and the
  effective observation time and is directly related to the total number
  of events recorded by a given instrument.}%
~exceeding $10^8$~m$^2$~sr~s, as indicated by the diagonal gray lines.

\section{Direct CR measurements with {\it Fermi}}

High-energy cosmic-ray electrons and positrons constitute a peculiar component
of the cosmic radiation in that, unlike protons and heavier nuclei, they
rapidly lose energy by synchrotron radiation on Galactic magnetic fields
and by inverse Compton (IC) scattering on the interstellar radiation field.
In the energy range of interest for the LAT---i.e. above a few GeV, where
losses dominate over escape---the average lifetime for electrons against
synchrotron and IC losses is $\tau_e(E) \propto E^{-1}$, whereas the corresponding
escape time for protons scales with the inverse of the diffusion coefficient:
$\tau_p(E) \propto E^{-\delta}$ (with $\delta \approx 0.3$--$0.6$).
This implies that the all-electron spectrum at Earth is significantly steeper
than the proton spectrum. More importantly, at energies above a few hundred
 GeV the majority of the electron and positron flux must be contributed by
sources closer than a few hundred pc.
In other words, high-energy CRE really probe cosmic-ray production and
propagation in nearby or local Galactic space.

The other piece of conventional wisdom is that unlike electrons, which are
accelerated in CR sources, the vast majority of positrons arise as secondary
products of cosmic-ray interactions in our Galaxy. In the absence of primary
sources of positrons it is straightforward to demonstrate that in the simplest 
scenarios the positron fraction is expected to decrease with energy like
$E^{-\delta}$, e.g. \cite{1998Moskalenko}.  As we shall see in a moment, this expectation is at odds with the most
recent measurements~\cite{PamelaPositrons}.

{\it Fermi} began its operations at about the same time two important
measurements~\cite{ATICElectrons,PamelaPositrons} were published that challenge
the simplest models of cosmic-ray production and propagation. These reports stirred 
an enormous excitement in the community, mainly in connection with the
possibility of detecting signs of new physics in the context of indirect dark
matter searches.
In 2009 the {\it Fermi} LAT collaboration published the first systematics-limited
measurement of the inclusive spectrum of cosmic-ray electrons and
positrons~\cite{CREFirstSpectrum} between 20~GeV and 1~TeV. Though harder than
previously thought, the {\it Fermi} spectrum does not show evidence of any prominent
feature, such as the one reported by the ATIC experiment~\cite{ATICElectrons}.
This conclusion was later independently confirmed by the
H.E.S.S.~\cite{HESSElectronsLE} and Pamela~\cite{PamelaElectrons} experiments,
as well as by a dedicated analysis of the {\it Fermi} data aimed at the highest end
of the energy range~\cite{CREFullPaper}. Fig. 2 summarizes the current status of measurements. 

\begin{figure}[htbp!]
  \begin{center}
    \includegraphics[width=\textwidth]{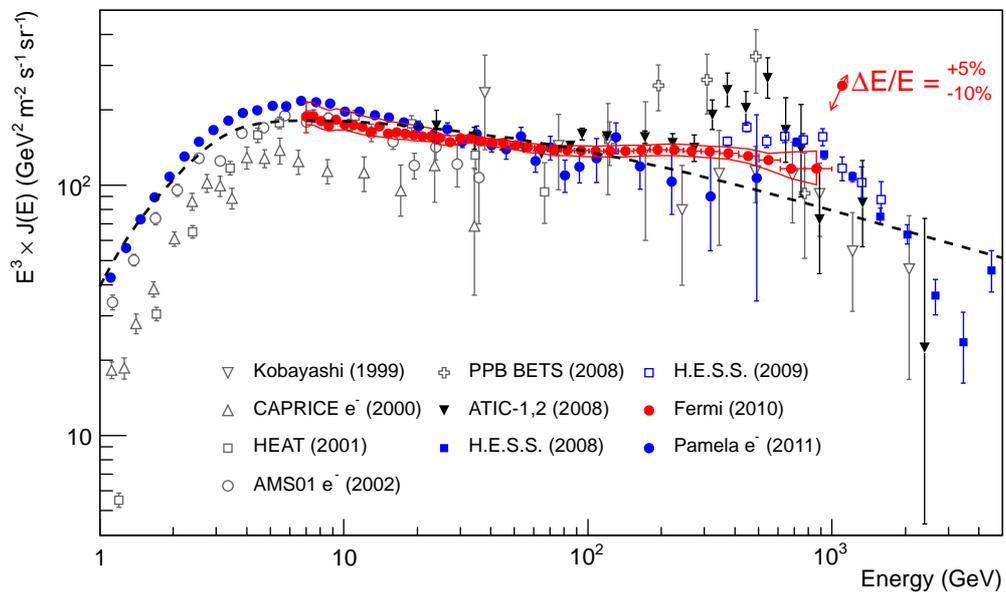}
  \end{center}
  \caption{Cosmic-ray electron plus positron spectrum measured by {\it Fermi}
    in one year of observation, adapted from~\cite{CREFullPaper}.
    Some of the most recent measurements are shown for comparison.
    The dashed line is a pre-{\it Fermi} model, shown for reference. Many of the data points are taken from the cosmic-ray database maintained by A.~W.~Strong and I.~V.~Moskalenko~\cite{GCRDB}.
    }
  \label{fig:cre_spectrum}
\end{figure}

The {\it Fermi} measurement---especially when the low-energy extension of the spectrum
published in~\cite{CREFullPaper} is taken into account---is in tension with
a single power-law model, though this cannot be definitely ruled out within the
systematic uncertainties. This tension has motivated a number of
papers in which a consistent interpretation of the {\it Fermi} inclusive spectrum
and the rise in the positron fraction above $\approx10$~GeV measured by
Pamela~\cite{PamelaPositrons} is attempted through the introduction
of an additional nearby source of high-energy electrons and positrons (for a
more detailed discussion see~\cite{CREInterpretation} and references therein).
It is fair to say that, to date, the issue is not settled, and there is no
firm consensus on whether observations really indicate the existence or nature of this
extra component.
Indeed there are alternative scenarios (see for
instance~\cite{BlasiPositrons,WaxmanAnomalies}) that might explain the
observations without invoking any new (astrophysical or exotic) source
of high-energy electrons and positrons.

For the future, with more than 5000 electrons above 1~TeV per year crossing the LAT,
{\it Fermi} can hope to provide an independent confirmation of the spectral cutoff
measured by H.E.S.S.~\cite{HESSElectrons} in this energy range. Even with a selection
efficiency on the order of 10\% (in line with the figures of merit of the
current analysis), the statistics will not be an issue.  At the same time, the collaboration is
actively working on the energy measurement, which is the most
challenging part of the analysis.

Without an onboard magnet, {\it Fermi} LAT cannot directly distinguish the
charge sign. However the Earth's magnetic field creates natural
\emph{shadow regions} from which particular charges are forbidden---simply
because their paths are blocked by the Earth---and can be effectively
exploited in order to separate positrons from electrons. The idea of
using the geomagnetic field to study the charge asymmetry in the leptonic
component of cosmic rays is not new. In fact the possibility of using  the east-west asymmetry in the geomagnetic cutoff rigidity has been explored
in several balloon experiments (see~\cite{GeomagPositrons} and references
therein for more information).
In addition to its large field of view and its excellent directional
capabilities, {\it Fermi} can take advantage of the accurate models of the
Earth's magnetic field~\cite{IGRF} and the particle tracing computer programs
that are nowadays readily available. One of the main challenges in the
analysis is the fact that, as the energy increases, the shadow regions become
smaller and smaller.  Coupled with the falling spectrum, this effect in practice limits the highest accessible
energy to $\approx200$~GeV. In addition to that, these regions are located near the Earth's horizon, which is outside or at the limits of the LAT
field of view in the nominal sky-survey operations.
Preliminary results of this analysis  are in
agreement with the positron fraction measured by Pamela and will be the
subject of a forthcoming paper. With the AMS detector now installed on the International
Space Station, new data in the unexplored high-energy range will be
available in the near future.

The electron component of cosmic rays is not fully described by
the inclusive energy spectrum and the positron fraction:
the degree of anisotropy in the arrival directions is an additional key
observable, especially in a scenario in which the inclusive electron spectrum
is dominated by one (or a few) nearby sources.
Some degree of anisotropy is indeed expected even in a plain diffusive
scenario in which the sources are continuously distributed, just due to the fact
that the Earth is not at the center of the diffusion halo. It is natural to
expect that the flux anisotropy can be much larger
if a single (or a few) nearby sources dominate the CRE spectrum
at high energy.

\begin{figure}[htbp!]
  \begin{center}
    \includegraphics[width=0.75\textwidth]{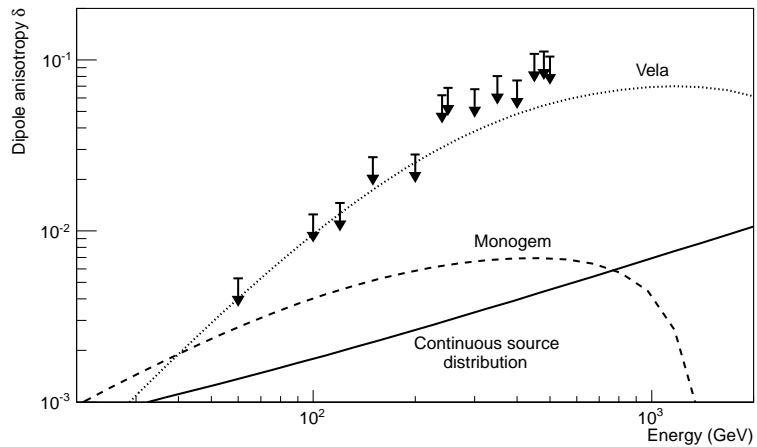}
  \end{center}
  \caption{{\it Fermi} LAT 95\% CL upper limits for the integral dipole anisotropy
    $\delta$ as a function of the minimum energy (adapted from~\cite{CREAnisotropies}).
    For comparison the expected anisotropy levels for two pulsars (Vela and
    Monogem) that are plausible candidate sources of high-energy electrons
    are shown. The values of the model parameters are described
    in~\cite{CREAnisotropies}; though subject to large uncertainties, the values are physically motivated and respect the relevant
    constraints (the first of which is the spectrum measured by {\it Fermi}).
    The (much lower) expected anisotropy for a continuous source distribution
    throughout the Galaxy (calculated with GALPROP \cite{1998Strong}) is also shown for reference.
    }
  \label{fig:cre_anisotropies}
\end{figure}

Key to these measurements are {\it Fermi}'s large geometric factor, its
excellent directional capabilities and its characteristic of an all-sky monitor
providing uniform exposure%
\footnote{The exposure is not actually uniform over the sky; there is a
  non-uniformity at the level of 15\%, mainly due to the fact that the
  instrument is not taking data while inside the South Atlantic Anomaly.
  This is described in~\cite{CREAnisotropies}.}%
.

{\it Fermi} has published the most stringent upper limits to the anisotropy
of the CRE flux, based on the first year of data~\cite{CREAnisotropies}.
The paper discusses all aspects of the analysis:
the pixelization of the sky, the pixel-to-pixel fluctuations, the search for
anisotropies on different angular scales (from $10^\circ$ to $90^\circ$) and
a full spherical harmonic analysis.
In this review we concentrate on a single aspect, namely the upper limits
on a possible dipole anisotropy, which is physically relevant for a single
source dominating the spectrum (in which case the source defines one of the
poles of the dipole).

Some of the basic results are shown in Figure~\ref{fig:cre_anisotropies}.
Though entering a detailed discussion of specific models and related
parameter space is beyond the scope of this review, we emphasize that {\it Fermi}
is already starting to constrain physically motivated scenarios and
might hope to detect a positive signal during the lifetime of the
mission (5 to 10 years). If indeed there is a nearby source of high-energy
electron/positron pairs, this signal would be a powerful tool to shed
light on its nature.  For most of the dark matter scenarios one
would expect an excess toward the Galactic center (potentially at the same
level as the ones shown in the figure) while Monogem and Geminga, which 
are two good pulsar candidates, are roughly in the opposite direction.

\section{Diffuse Galactic Gamma Rays as Probes of Galactic Cosmic Rays}

\begin{figure}[t]
\vspace*{2mm}
\begin{center}
\includegraphics[width=0.5\textwidth]{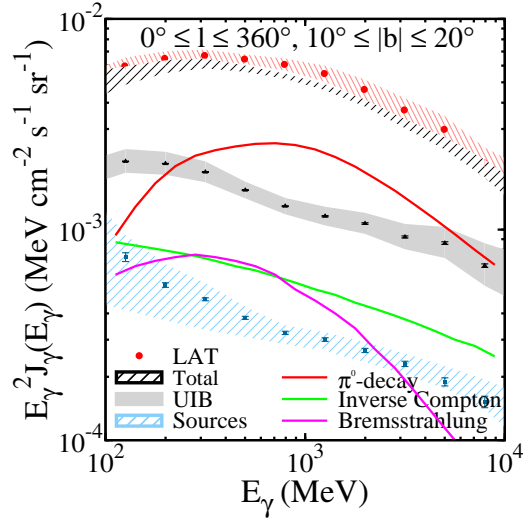}
\end{center}
\caption{Spectral Energy Distribution of diffuse $\gamma$ radiation at intermediate Galactic latitudes. Overlaid are SEDs for the component processes of the diffuse emission. The UIB component is unidentified isotropic background, including instrumental background.  The primary component, shown by the heavier line above the gray band, is decay of $\pi^0$ mesons~\cite{LAT09_GeVexcess}.}
\label{fig:SED_noGeV}
\end{figure}

Unlike the sky at visible wavelengths, the $\gamma$-ray sky is strongly dominated by diffuse radiation originating in our Milky Way Galaxy, making the Galactic plane the most striking feature in any large-scale $\gamma$-ray skymap.  Although the diffuse Galactic $\gamma$ radiation is brightest along the plane, it is seen at all Galactic latitudes.  This radiation is largely produced by cosmic-ray interactions with the interstellar gas and photon fields through the processes of inelastic nucleon scattering, bremsstrahlung, and inverse Compton scattering.   Figure~\ref{fig:SED_noGeV} shows the Spectral Energy Distribution (SED) of this radiation in part of the sky, along with modeled components~\cite{LAT09_GeVexcess}.  The largest contributor to this radiation is nucleon-nucleon collisions (basically cosmic ray protons and heavier nuclei hitting interstellar hydrogen nuclei) with subsequent decay of $\pi^0$ mesons into $\gamma$ rays in the {\it Fermi} LAT energy range.   The LAT measurements do not confirm the unexpected excess in the few GeV energy range measured by EGRET on the {\it Compton Gamma Ray Observatory (CGRO)} compared to models. 

\begin{figure}[t]
\vspace*{2mm}
\begin{center}
\includegraphics[width=0.75\textwidth]{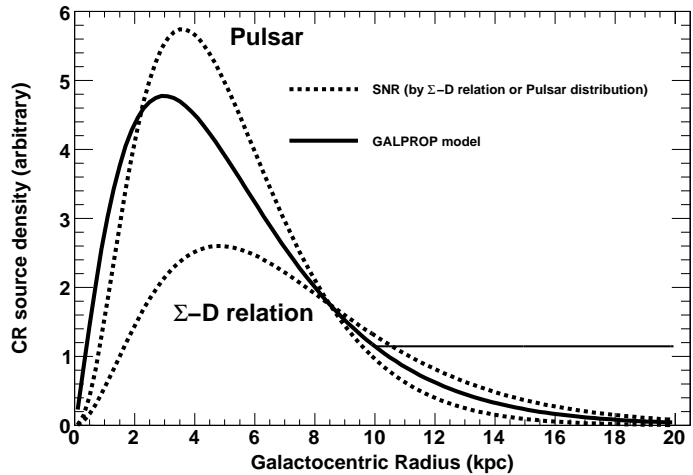}
\end{center}
\caption{CR source distribution from one GALPROP model (solid line), compared with the SNR distribution obtained by the $\Sigma$$-$D relation~\citep{1998Case} (see \citep{2009Green} for a critique of this method, however) and that traced by the pulsar distribution~\citep{2004Lorimer} shown by dotted lines. The thin solid line represents an example of the modified distributions introduced to reproduce the emissivity gradient observed by the LAT~\cite{2011_LAT_CR_Outer}.}
\label{fig:CRsource}
\end{figure}

Mapping the Galactic $\gamma$ rays offers a way to derive information about the spatial distribution of cosmic-ray particles.  The modeling necessary to extract such results is not simple because it requires knowledge of all the matter and photon distributions with which the cosmic rays interact.  Models for distribution of cosmic-ray sources and propagation include GALPROP~\citep{GALPROP04}\footnote{see also {\it http://galprop.stanford.edu/}} and a derivative, DRAGON~\citep{DRAGON}.  Figure~\ref{fig:CRsource} shows one result that has been derived from the LAT results:  an indication that conventional models for CR source distributions underpredict the observations in the outer Galaxy.  As noted in~\citet{2011_LAT_CR_Outer}, the observed $\gamma$-ray emissivity gradient is inconsistent with all available tracers of massive star formation.  In this case the LAT results suggest either a large scale height for the CR diffusive halo or a flatter distribution of the sources beyond the Solar circle, although other possibilities cannot be ruled out, including a large mass of unseen gas in the outer Galaxy~\citep{Papadopoulos}, a non-uniform diffusion coefficient~\citep {DRAGON}, or advective CR transport by a Galactic wind~\citep{Breitschwerdt}. 

\begin{figure}[t]
\vspace*{2mm}
\begin{center}
\includegraphics[width=0.75\textwidth]{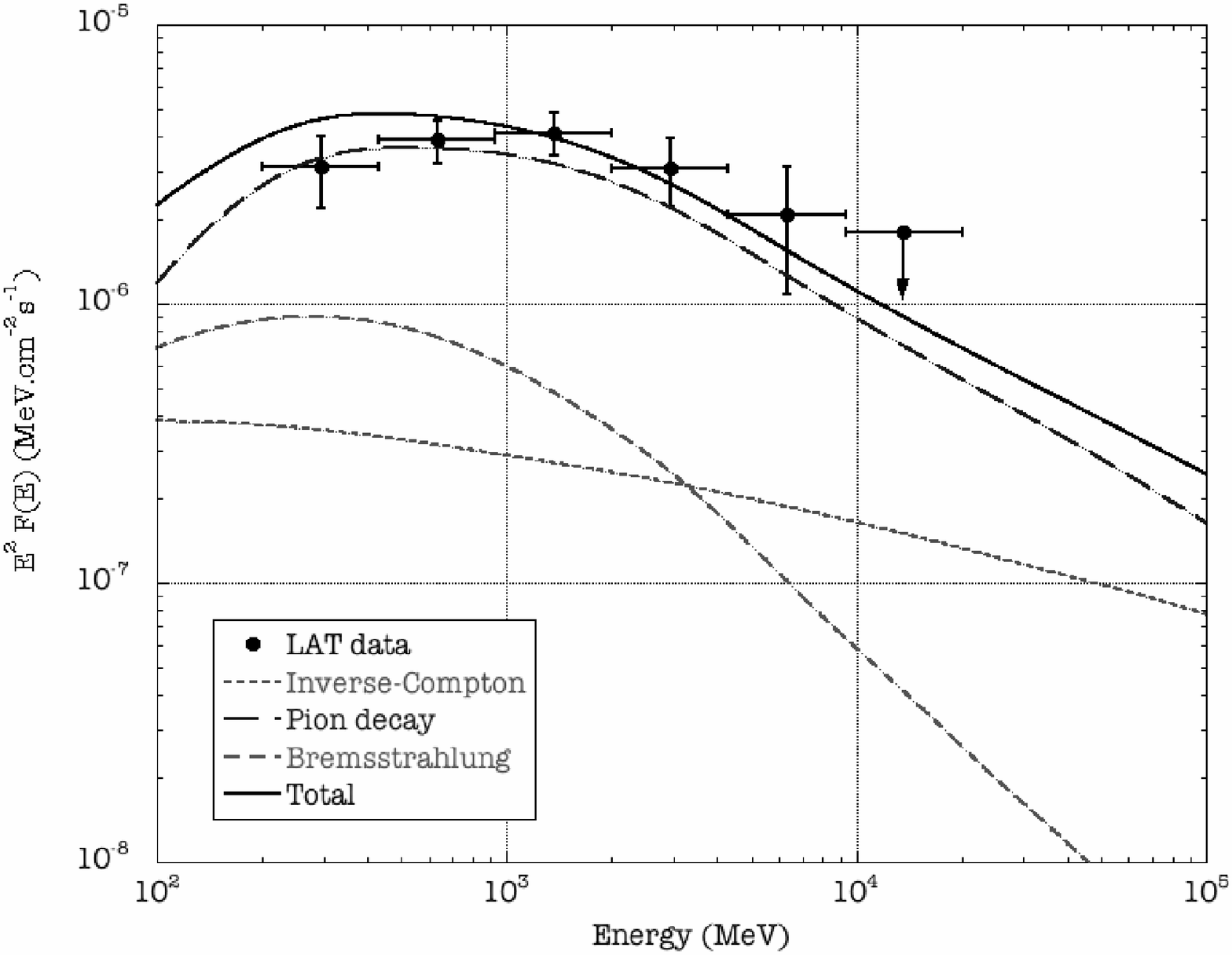}
\end{center}
\caption{SED of $\gamma$ radiation from the Small Magellanic Cloud. Overlaid are SEDs for the component processes of the diffuse emission. The primary component is decay of $\pi^0$ mesons~\cite{LAT10_SMC}.}
\label{fig:SED_SMC}
\end{figure}

Some galaxies not exhibiting apparent nuclear jet activities are also seen as $\gamma$-ray sources that offer information about cosmic rays beyond our own Galaxy.  Examples are normal galaxies, including the Large Magellanic Cloud~\cite{LAT10_LMC}, the Small Magellanic Cloud~\cite{LAT10_SMC}, and starburst galaxies, such as M82 and NGC 253~\cite{LAT10_starbursts}.  GeV $\gamma$ rays in these galaxies come primarily from the interactions of cosmic ray hadrons and electrons with interstellar matter and photon fields.  Figure~\ref{fig:SED_SMC} shows the measured spectrum from the Small Magellanic Cloud.  The spectrum is modeled predominantly by $\pi^0$ decay, implying hadronic processes similar to those in our own Galaxy.  The detailed analysis of the LMC shows that the CR flux there is less than a third that of the Milky Way~\cite{LAT10_LMC}, and the derived CR flux in the SMC is even lower~\cite{LAT10_SMC}. 
The $\gamma$-ray luminosities of these normal and starburst galaxies show an approximately linear relationship with the product of the supernova rate and the total mass of gas in the galaxies~\cite{LAT10_starbursts}.  The suggestion is that the $\gamma$-ray production requires both target material (gas) and CR production (traced by the supernova rate).  Although this analysis is highly simplified, the trend supports the expectation that the principal accelerators of cosmic rays are in some way related to massive-star-formation activity.

\section{Supernova Remnants and Other Candidate Cosmic-Ray Sources}

\subsection{SNRs as Cosmic-Ray Sources}

Supernova remnants (SNRs) have long been thought to be the main sources of 
Galactic CRs~\cite{Ginzburg64}, accelerating CR particles 
at their expanding shock waves through the diffusive shock acceleration (DSA) 
process \cite{BE87}. 
In this standard paradigm, SNRs are responsible for the production of 
cosmic-ray protons and nuclei up to energies of $\sim 3\times 10^{15}$ eV, 
transferring $\sim 10\%$ of the explosion kinetic energy into the form of 
cosmic-ray energy \cite{Hillas_Review05}. 
The observations of $\gamma$ rays from SNRs have been viewed as the most 
promising method of addressing the SNR paradigm for the origin of cosmic rays, 
through 
the measurements of the $\pi^0$-decay (or \emph{hadronic}) $\gamma$ rays \cite{DAV94}. 
 Despite the substantial progress of TeV $\gamma$-ray observations 
in recent years \cite{2009_TeVREview}, 
 finding a convincing case of $\pi^0$-decay $\gamma$ rays 
 remained tantalizingly difficult, mainly because electromagnetic 
 radiation processes involving relativistic electrons (the so-called \emph{leptonic} 
 components) could explain the $\gamma$-ray emission as well. 
  It has widely 
 been expected that the \emph{Fermi} LAT could identify the 
 hadronic component of the $\gamma$-ray emission in SNRs. 
Indeed the observations with the {\it Fermi} LAT have started to provide 
an effective means to disentangle the leptonic (IC scattering and 
relativistic bremsstrahlung) 
and hadronic ($\pi^0$-decay) $\gamma$-ray components. 

\begin{figure}
\includegraphics[width=0.7\textwidth]{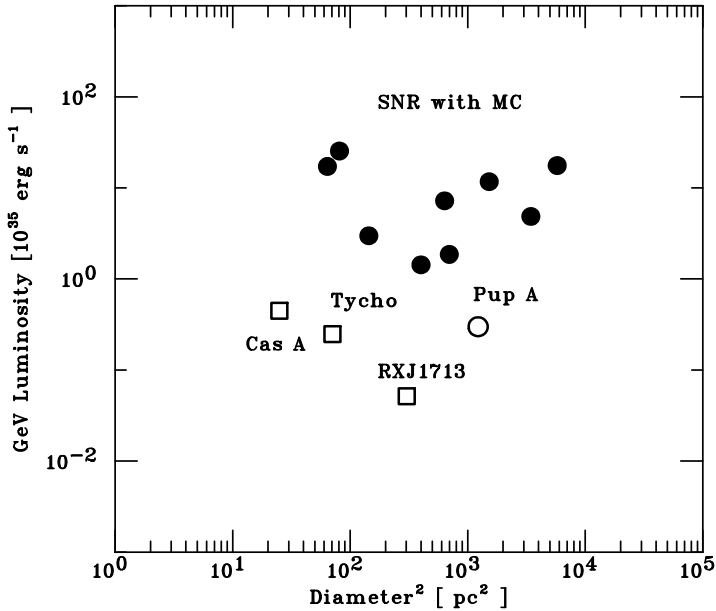}
\caption{
 GeV luminosity (0.1-100 GeV) in units of $10^{35}\ \rm erg\ s^{-1}$ 
 plotted as a function of diameter squared for the SNRs detected 
 with the \emph{Fermi} LAT. Filled circles correspond to 
 the SNRs interacting with molecular clouds. Rectangles are young 
 SNRs. Pup~A, middle-aged without clear molecular interactions, 
 is shown as an open circle.  }
\label{fig:SNR_Summary}
\end{figure}

Gamma-ray observations with 
the \emph{Fermi} LAT  have so far resulted in the detections of 
GeV $\gamma$-ray emission from more than ten SNRs \cite{W51C,W44,IC443,W28,W49B,Castro,CasA,RXJ1713}. 
 Figure~\ref{fig:SNR_Summary} summarizes  the observed GeV luminosity 
in 0.1--100 GeV as a function of the diameter squared for 
 the known LAT-detected SNRs. 
 Note that the squared diameter would be regarded as a reasonable
  indicator of SNR ages, as 
the age of a remnant in the Sedov-Taylor phase is related to 
its radius as $t \propto R^{2.5}$.  
The remnants interacting with molecular clouds are indicated by filled circles.
They constitute the dominant class of the LAT-detected SNRs, 
and they are generally more luminous in the GeV band than young SNRs~\cite{Uchi11}. 
The observed $\gamma$-ray luminosity of the SNRs interacting with molecular clouds is so high that one of the main leptonic radiation channels, namely 
IC scattering off the Cosmic Microwave Background (CMB) and interstellar radiation fields, becomes 
unlikely. 
For an efficiency of $\sim 10\%$ for CR proton production and 
an electron-to-proton ratio of 0.01, the GeV luminosity of the IC emission 
is calculated to be $L_{\rm IC} \sim 10^{34}\ \rm erg\ s^{-1}$ using the 
 interstellar radiation fields of the solar neighborhood. 
The observed luminosity well exceeds this value as shown in 
Figure~\ref{fig:SNR_Summary}.
In what follows, we review the results obtained with the LAT in the context of 
the SNR paradigm for CR origin. 

\subsection{Young Supernova Remnants} \label{sec:YoungSNRs}

The remnants of historical supernovae are
generally well studied across the electromagnetic spectrum,
 making them  excellent laboratories for studying 
high-energy phenomena associated with supernova shocks, and 
for testing our current understanding of the DSA theory. 
The following issues can be addressed by the \emph{Fermi} LAT 
observations of young SNRs: 

(i) {\it Acceleration efficiency}---About 10\% of the mechanical energy of supernova explosions must be transferred to the kinetic energy of CR protons in order to 
maintain the density of Galactic CRs. 
Since the detailed mechanisms of 
how thermal particles are injected into the acceleration process 
and therefore the efficiency of particle injection remain uncertain, 
the current theory does not firmly predict the amount of CR protons 
in supernova shocks. The amount of CR electrons produced at shocks is even  
more uncertain theoretically; 

(ii) {\it Magnetic field amplification}---
The DSA process at supernova shocks depends strongly on the generation of 
magnetic field turbulence by CR themselves ahead of the shock wave. 
It has been predicted that 
the turbulent magnetic field in the strong shocks of young SNRs can be 
significantly amplified  by CR current-driven instability \cite{Bell04} 
and other instabilities. 
Evidence for the magnetic field amplification was provided by 
high angular resolution observations with the \emph{Chandra} X-ray 
Observatory. Narrow widths of synchrotron X-ray filaments \cite{VL03,Bamba03} 
and their short time variability \cite{Uchi07} require strong magnetic fields. 
Acceleration of protons up to the knee in the CR spectrum at $\sim 10^{15}$ eV 
can be realized only if magnetic field amplification occurs. 
Gamma-ray observations can put an important constraint on 
the strength of magnetic field, since the $\gamma$-ray flux due to 
IC scattering is scaled simply by  
magnetic field strength given the observed flux of synchrotron radio emission; 

(iii) {\it Acceleration spectrum}---
In the test-particle regime, the DSA theory predicts the spectrum of CRs 
accelerated at SNR shocks obeying a power law $Q(E) \propto E^{-s}$ 
with $s \simeq 2$. 
The CR spectra measured directly at Earth 
have the form $N(E)\propto E^{-2.7}$ below the knee energy, 
as confirmed by the \emph{Fermi} LAT measurements of 
the Galactic diffuse $\gamma$-ray emission. 
Then the diffusion coefficient for  CR propagation should be roughly 
$D(E) \propto Q(E)/N(E) \propto E^{0.7}$. 
Such a strong energy dependence would be inconsistent with the 
CR anisotropy measurements \cite{Ptuskin06,BlasiAmato11}.
Recently it has been suggested that the finite velocity of scattering centers 
(usually neglected in the DSA theory) 
due to magnetic field amplification may lead to 
steeper spectra of accelerated CRs with $s \sim 2.2$ \cite{Caprioli10}, 
though the prediction  depends on poorly understood properties of 
the magnetic turbulences produced at shocks. 
In this regard, 
measurements of the GeV $\gamma$-ray spectrum, especially that of 
hadronic origin, are of great importance.

\subsubsection{Cas A}

Cas A is thought to be the remnant of a Type IIb supernova that 
exploded around A.D. 1680. 
Since it is at the early Sedov-Taylor phase,  
the accelerated CR spectrum can reach already 
its maximum attainable energy and the process of energy 
conversion, from the expansion kinetic energy into CR energy, 
can well proceed. 
Cas A was the first SNR detected in the TeV $\gamma$-ray 
band \cite{HEGRA_CasA}.
Though the emission mechanisms responsible for the TeV $\gamma$ rays 
remained unsettled, their detections
provided direct evidence for particle acceleration beyond 10 TeV.
Observations with EGRET resulted in setting upper limits on the $\gamma$-ray 
flux above 100 MeV. 

\begin{figure}
\begin{center}
\includegraphics*[width=0.7\textwidth]{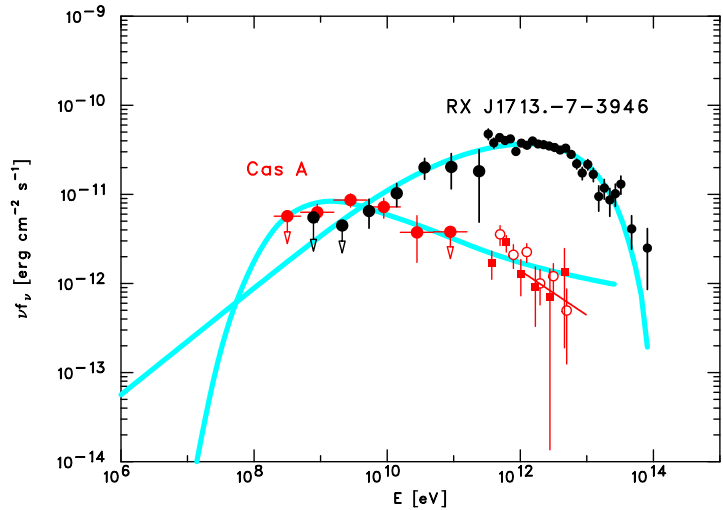}
\end{center}
\caption{Gamma-ray spectra of young SNRs: Cas~A (red)
and RX~J1713.7$-$3946 (black). 
Cas~A: the LAT spectrum in 0.5--50 GeV (filled circles)~\cite{CasA} is shown 
along with the TeV spectra reported by 
HEGRA (line)~\cite{HEGRA_CasA}, 
MAGIC (squares)~\cite{MAGIC_CasA}, and VERITAS (open circles) 
~\cite{VERITAS_CasA}. 
RX~J1713.7$-$3946: the LAT spectrum 
in 0.5--400 GeV (filled circles)~\cite{RXJ1713} is shown 
in combination with the energy spectrum of 
H.E.S.S.-detected emission (small filled circles)~\cite{HESS_1713_2}.  
The $\gamma$-ray spectrum of Cas~A is best described by 
a hadronic emission model ($\pi^0$-decay $\gamma$ rays), while 
the $\gamma$-ray emission from RX~J1713.7$-$3946 can be fitted 
by a leptonic model (inverse-Compton scattering). 
 }
\label{fig:YoungSNRs_SED}
\end{figure}

A clear detection of Cas~A at a significance level of $12\sigma$ was made 
with the {\it Fermi} LAT using  the $\gamma$-ray data 
during the period between 2008 August 4 and 2009 September 4 \cite{CasA}. 
The GeV $\gamma$-ray data were consistent with a point-like source, 
as expected from the angular size of the remnant (a radius of $2.5^\prime$), and also 
consistent with steady emission. 
The $\gamma$-ray spectrum measured for Cas~A 
in the GeV--TeV band is shown in Figure~\ref{fig:YoungSNRs_SED}.

The X-ray observations of synchrotron-emitting filaments measure the magnetic field at the forward shock; 
$B \simeq 0.3\ \rm mG$ is inferred from the observed width of the filaments 
\cite{Parizot06}, which is consistent with the variability timescale of 
the synchrotron filaments~\cite{UA08}. 
By combining the estimated magnetic field 
 with the synchrotron spectrum in the radio bands, 
it was shown that  the leptonic $\gamma$-ray emission model 
 does not well explain the LAT-detected emission \cite{CasA}. 
Instead, the GeV--TeV $\gamma$-ray emission can be ascribed mainly to 
the $\pi^0$-decay component. The total proton content 
at the current age of the remnant 
amounts to 
$W_p \simeq 0.4\times 10^{50}\ \rm erg$ for the 
shocked gas density of $n_{\rm H}= 10\ \rm cm^{-3}$, which 
is estimated from  the measured dynamics of the remnant. 
This would be the first case where one can estimate the 
amount of CRs in SNRs with a reasonable accuracy. 
The spectral slope of accelerated CR protons is assumed to be 
$s = 2.3$ from the radio spectral index (Figure~\ref{fig:YoungSNRs_SED}), 
which seems compatible with the GeV--TeV spectrum. 
Both the CR content and spectrum are in reasonable agreement with 
the direct measurements  of the CR flux and the CR anisotropy.

\subsubsection{Tycho}

Tycho's SNR is the remnant of a Type Ia explosion in 1572,  
currently in transition from the ejacta-dominated phase to 
the Sedov-Taylor phase. 
Recent detection of  TeV $\gamma$-rays from Tycho's SNR 
by the VERITAS Collaboration \cite{VERITAS_Tycho} has just been 
followed by the very recent $5\sigma$ detection of GeV $\gamma$-ray emission 
with the \emph{Fermi} LAT~ \cite {LAT_Tycho}. 
The LAT spectrum is characterized by photon index 
$\Gamma =2.3 \pm0.2$(stat)$\pm 0.1$(sys), which smoothly connects with 
the VERITAS spectrum. 

Given the amplified magnetic field of 
0.2--0.3 mG estimated for shock downstream \cite{Voelk05,Parizot06} 
and the constraint on the ambient density ($n_{\rm H} < 0.3\ \rm cm^{-3}$), 
the $\gamma$-ray spectrum of Tycho's SNR can be explained only by 
the $\pi^0$-decay emission. 
Compared with Cas~A, 
 the ambient density is smaller and the radio emission is much weaker, 
 so that the leptonic scenario faces a more serious difficulty. 
The total CR proton content at the current age 
amounts to 
$W_p \simeq (0.6\mbox{--}1.5)\times 10^{50}\ \rm erg$ for a 
reasonable range of SNR parameters such as distance. 
The accelerated CR protons and electrons with number index $s=2.3$ 
are inferred from the multiwavelength fitting. 
Such a steep spectrum may be explained by the finite velocity of 
the scattering waves in the shock upstream region \cite{MorlinoCaprioli11}.
As for  Cas~A, 
the LAT results obtained for Tycho's SNR strengthen the case for SNR origin of 
the Galactic CRs.

\subsubsection{RX~J1713.7$-$3946}

SNR RX~J1713.7$-$3946 (G347.3$-$0.5 in the Green's SNR list~\cite{Green})
is well known for its bright synchrotron X-ray emission that completely dominates 
the radiation output from the remnant~\cite{Koyama97,Slane99} 
and also for its bright TeV $\gamma$-ray emission 
that well traces the synchrotron X-ray map 
~\cite{HESS_1713,HESS_1713_2}.
Since its discovery, 
the origin of the TeV $\gamma$-ray emission has attracted strong 
interest from the high-energy astrophysics community \cite{Zira10,Ellison10}.

\begin{figure}
\begin{center}
\includegraphics*[width=11cm]{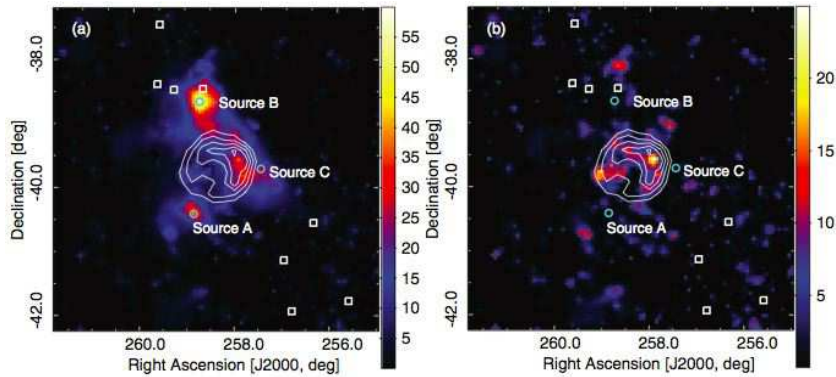}
\end{center}
\caption{Test Statistic maps for a point source hypothesis at each 
location of the sky in the region around SNR RX~J1713.7$-$3946~\cite{RXJ1713}.
Events above 0.5 GeV are analyzed. TS maps were obtained by 
maximum likelihood fitting, where (a) the Galactic and isotropic diffuse emissions 
and 1FGL catalog sources \citep{LAT10_1FGL} were included as background sources, (b) 
Sources A, B, and C were also taken into account in the background model.}
\label{fig:RXJ1713_map}
\end{figure}

Figure \ref{fig:RXJ1713_map} presents the Test Statistic (TS) map of 
a sky region around SNR RX~J1713.7$-$3946 calculated 
using the events above 0.5 GeV taken from 
2008 August 4 to 2010 August 4~\cite{RXJ1713}. 
(TS is defined as $\mbox{TS} = -2\ln (L_0/L_{\rm point})$ where $L_0$ 
is the likelihood value of the null hypothesis that takes account of 
only background sources in the model, and  $L_{\rm point}$ is 
that of the hypothesis assuming a point source at each location.)
The extended $\gamma$-ray emission coincident with SNR RX~J1713.7$-$3946 
can be seen in Figure \ref{fig:RXJ1713_map}, though it was hard to 
investigate internal spatial structures due to the lack of photons. 

In Figure~\ref{fig:YoungSNRs_SED}, the energy spectrum of 
SNR RX~J1713.7$-$3946 measured with the \emph{Fermi} LAT \cite{RXJ1713} 
is shown together with that of the H.E.S.S. telescopes \cite{HESS_1713_2}. 
The LAT spectrum can be
characterized by a hard power law with photon index 
$\Gamma =1.5 \pm0.1$(stat)$\pm 0.1$(sys), 
smoothly connecting with the steeper TeV spectrum. 
The hard power-law shape agrees with the
expected IC spectrum (the leptonic model), as illustrated in 
Figure~\ref{fig:YoungSNRs_SED}. 
Given the energy flux ratio of 
the observed synchrotron X-ray emission and the $\gamma$-ray 
emission,  the leptonic model requires that the average magnetic field be 
weak, $B \simeq 10\,\mu$G \cite{HESS_1713_2}. 
If the leptonic scenario is the case, 
the filamentary structures and
variability in X-rays \cite{Uchi07} would be attributable to locally enhanced
magnetic fields, e.g., time-variable local turbulent magnetic fields \cite{Bykov09}.
Importantly, 
within the leptonic scenario the $\gamma$-ray spectrum provides a robust estimate of 
the total amount of relativistic electrons, as $W_e \sim 1\times 10^{48}\ \rm erg$.

\subsection{SNRs Interacting with Molecular Clouds} \label{sec:MC-SNRs}

\begin{figure}
\begin{center}
\includegraphics*[width=8cm]{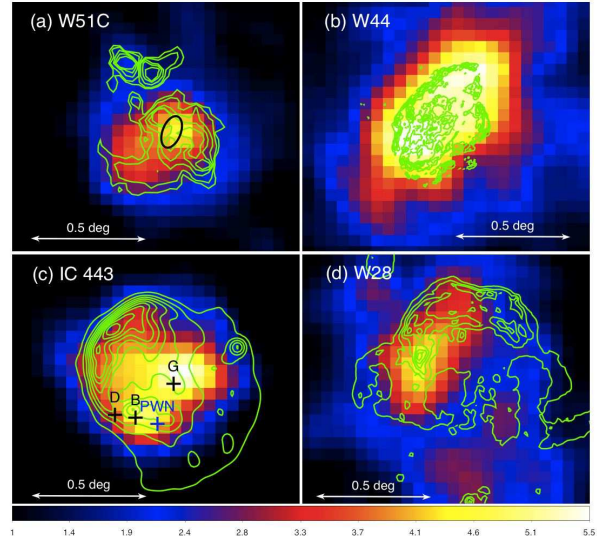}
\end{center}
\caption{
\emph{Fermi} LAT 2--10 GeV count maps (in celestial coordinates)  of 
 the MC-interacting SNRs with extended 
 $\gamma$-ray emission: (a) W51C; (b) W44; (c) IC443; and (d) W28 
~\cite{Uchi11}.
The intensity 
is  in units of counts per pixel with a pixel size of $0.05\times 0.05$~deg$^2$. 
Superposed are the contours from the VLA radio maps (see~\cite{Uchi11} 
and references therein). 
A black ellipse in panel (a) 
represents the location of shocked CO clumps.
The black crosses in panel (c) are the locations of  shocked molecular clumps 
from which OH maser emission is detected. The position of 
a pulsar wind nebula is also marked in panel (c). 
}
\label{fig:SNR_images}
\end{figure}

The brightest GeV $\gamma$-ray sources associated with SNRs 
are middle-aged remnants that are interacting with molecular clouds (MCs) 
as evidenced by 
the detections of  1720 MHz OH maser emission \cite{Hewitt09}. 
Figure~\ref{fig:SNR_images} presents 
the raw LAT count maps in 2--10 GeV of the four SNRs 
(W51C \cite{W51C}, W44 \cite{W44}, 
 IC~443 \cite{IC443}, and W28 \cite{W28})
obtained with 2.5 yr data of the LAT observations \cite{Uchi11}.
Even before the subtraction of the intense Galactic diffuse emission, 
the $\gamma$-ray signals from the SNRs  can be easily seen.

\begin{figure}
\begin{center}
\includegraphics*[width=8cm]{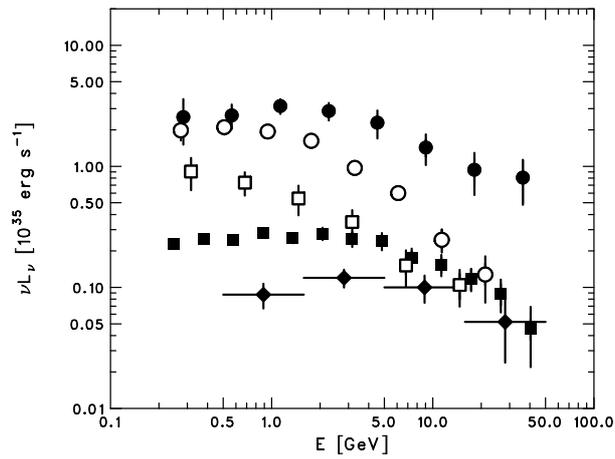}
\end{center}
\caption{
Gamma-ray spectra of shell-type SNRs measured 
with the \emph{Fermi} LAT: W51C (filled circles~\cite{W51C}); 
W44 (open circles~\cite{W44}); 
IC~443 (filled rectangles~\cite{IC443}); 
W28 (open rectangles~\cite{W28}); 
Cassiopeia~A (filled diamonds~\cite{CasA}).  
}
\label{fig:SNR_spectra}
\end{figure}

The MC-SNRs (namely, 
SNRs interacting with MCs) constitutes the dominant class of LAT-detected 
SNRs (Figure~\ref{fig:SNR_Summary}).
The $\gamma$-ray  luminosity in the 1--10 GeV band 
spans 
$L_\gamma = (0.8\mbox{--}9)\times 10^{35}\, \rm erg\ s^{-1}$, 
larger than for the young SNRs like Cas~A and RX~J1713.7$-$3946 
(Figure~\ref{fig:SNR_spectra}). 
The GeV $\gamma$-ray spectra commonly exhibit 
a spectral steepening in the \emph{Fermi} LAT band, demonstrating 
the importance of LAT observations of these SNRs. 
Also, the synchrotron radio emission of the four SNRs can be characterized 
by a large flux density of 160--310 Jy at 1 GHz 
with flat spectral index of $\alpha \simeq 0.3\mbox{--}0.4$ \cite{Green}. 
The LAT-detected SNRs are generally radio-bright objects \cite{Uchi11}.

The predominance of the MC-SNRs among 
the SNRs detected by the \emph{Fermi} LAT and their 
high  $\gamma$-ray  luminosity 
indicate that the $\gamma$-ray emission should be enhanced by 
the interactions with molecular clouds. 
Nonthermal  bremsstrahlung by relativistic electrons 
or $\pi^0$-decay $\gamma$-rays produced by high-energy protons are 
the two plausible channels of the $\gamma$ radiation. 
As discussed above, 
the observed high luminosity of the GeV $\gamma$-ray emission disfavors 
the IC scattering off the CMB and interstellar radiation fields 
as the emission mechanism.

\begin{figure}
\begin{center}
\includegraphics*[width=9cm]{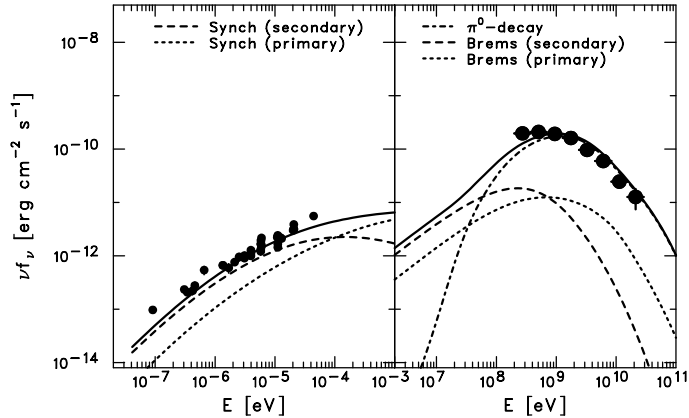}
\end{center}
\caption{
Radio (left) and $\gamma$-ray (right) spectra of SNR W44 together with 
the reacceleration model~\cite{Uchi10}. 
}
\label{fig:W44_SED}
\end{figure}

There are two different types of models to account for  the GeV $\gamma$-ray 
emission from the MC-SNR systems.
The {\it Runaway CR} model 
considers  $\gamma$-ray emission from 
molecular clouds illuminated by runaway CRs 
that have escaped from their accelerators, namely 
SNRs \citep{AA96,Gabici09,Ohira11}. 
If the $\gamma$-ray emission detected by the \emph{Fermi} LAT can be 
ascribed to the escaping CRs, 
one can learn about how CRs are released into interstellar space and 
how they propagate in the vicinity of the SNR where 
self-generated Alfv\'en waves may change the local diffusion 
coefficient \cite{Fujita10}.
Another scenario is the so-called {\it Crushed Cloud} model \citep{BC82,Uchi10} 
that invokes a  ``shocked" molecular cloud into which a radiative shock 
(typically $v_{\rm s} \sim 100\ \rm km\ s^{-1}$) 
is driven by the impact of the SNR's blastwave. 
The cosmic-ray particles accelerated at a cloud shock are 
adiabatically compressed behind the shock front, resulting in enhanced 
synchrotron and $\pi^0$-decay $\gamma$-ray emissions. 
As shown in Figure~\ref{fig:W44_SED}, 
reacceleration of 
pre-existing CRs at a cloud shock is capable of explaining both radio synchrotron 
and GeV $\gamma$-ray emission in the case of SNR W44 \cite{Uchi10}.
The $\gamma$-ray luminosity of  $L_\gamma \sim 10^{35}\, \rm erg\ s^{-1}$ 
in 1--10 GeV agrees with the theoretical expectation. 
Morphologies of synchrotron radio emission support this scenario. 
The radio continuum map of SNR W44 
exhibits filamentary and sheet-like structures  of synchrotron radiation 
well correlated with the shocked H$_2$ emission \cite{Reach05}. 
One can learn about shock-acceleration processes in dense environments, 
which complements the studies of young SNRs. 

Both sites of the $\gamma$-ray emission could be important. 
 For example, 
a hard GeV $\gamma$-ray source \cite{W28} 
which is also bright in the TeV band \cite{HESSW28} 
outside the southern boundary of SNR W28 might be explained by runaway CRs, 
while $\pi^0$-decay emission from 
radiatively-compressed layers of clouds could account for the GeV 
$\gamma$ rays observed within SNR W28. 
Detailed analysis of spatial distributions of GeV $\gamma$ rays 
combined with CO observations 
will  make it possible to discern the $\gamma$-ray sources that are attributable 
to runaway CRs. 

\subsection{Other Candidate Cosmic-Ray Sources}

Although SNR remain primary candidates for accelerating the bulk of cosmic rays, other Galactic sources observed by the {\it Fermi} LAT also represent sites of particle acceleration and interaction.  Such sources include 
pulsar wind nebulae \cite{LAT11_PWNcat}, high-mass binary systems such as  LS I $+61^\circ 303$~\citep{LAT09_LSI} and Cygnus X-3~\citep{LAT09_CygX3}, whose spectra seem most consistent with leptonic sources, and Nova V407 Cygni~\citep{LAT10_V407Cyg}, a symbiotic binary system whose spectrum can be modeled with either leptonic or hadronic processes. 
While the energetics requirement for the CR protons is generally difficult to 
satisfy by  these non-SNR sources, they might be the dominant sources of the CR electrons and positrons. The large number of unidentified sources found in the {\it Fermi} LAT 1FGL catalog~\citep{LAT10_1FGL} offers the possibility that there may be yet-undiscovered CR production sources.

\section{Conclusions}

Both electron and $\gamma$-ray observations made with the {\it Fermi} Large Area Telescope offer new insights into cosmic-ray acceleration and interactions.  The direct electron plus positron measurements suggest, although they do not prove, a possible nearby source.  Future LAT observations will help measure separately the positive and negative components up to 200 GeV and will improve on the search for anisotropies already begun.  On a Galactic scale,  $\gamma$ rays suggest a larger scale height or flatter radial distribution of cosmic rays than previously expected.  
From the $\gamma$-ray studies of supernova remnants, LAT has found growing evidence that at least some SNR are likely accelerators of hadronic cosmic rays, although one example, RX~J1713.7$-$3946, appears more consistent with being a leptonic source.  
The {\it Fermi} LAT continues to collect data.  Improved analysis methods, coupled with the higher statistics from the ongoing mission, can confidently be expected to produce even more results addressing questions about cosmic-ray origin and propagation in the future. 

\section{Acknowledgments}

The authors express their deepest thanks to the {\it Fermi} Large Area Telescope team for building this wonderful instrument and providing the opportunity for so many exciting results.  We particularly thank T. Brandt, I. Moskalenko, and O. Tibolla for their valuable suggestions during the preparation of this review. 





\bibliographystyle{elsarticle-num}
\bibliography{Cosmic_Rays_Review_LAT_vX}







\end{document}